      [1999/12/01 v1.4c Il Nuovo Cimento]
\documentclass{cimento}

\usepackage{graphicx}

\def\gtrsim{\lower.5ex\hbox{$\; \buildrel > \over\sim \;$}} 
\def\lesssim{\lower.5ex\hbox{$\; \buildrel < \over\sim \;$}}

\title{High-Energy Cosmic Rays and Neutrinos from Gamma-Ray Bursts}
\author{C.~Dermer\from{ins:x}}
\instlist{\inst{ins:x} U.\ S.\ Naval Research Laboratory, Washington, D.\ C.}
\PACSes{\PACSit{96.40.Tv, 98.70.Rz, 98.70.Sa}
\PACSit{96.40.Tv, 98.70.Rz, 98.70.Sa}}

\begin{document}

\maketitle

\begin{abstract}
A complete model for the origin of high-energy ($\gtrsim 10^{14}$ eV)
cosmic rays from gamma-ray bursts (GRBs) and implications of this
hypothesis are described. Detection of high-energy neutrinos from GRBs
provide an unambiguous test of the model. Evidence for cosmic-ray
acceleration in GRBs is suggested by the detection of anomalous
$\gamma$-ray components such as that observed from GRB 941017. Neutron
$\beta$-decay halos around star-forming galaxies such as the Milky Way
are formed as a consequence of this model. Cosmic rays from GRBs in
the Galaxy are unlikely to account for the $\sim 10^{18}$ eV
cosmic-ray excess reported by the Sydney University Giant Air Shower
Recorder (SUGAR), but could contribute to past extinction
events.
\end{abstract}

\section{Introduction}

One of the outstanding questions in contemporary astronomy is the
origin of HECRs ranging from below the knee of the cosmic-ray spectrum
at $\approx 3\times 10^{15}$ eV to the highest energies exceeding $
10^{20}$ eV.  Ultra-high energy cosmic rays (UHECRs), defined here as
cosmic rays (CRs) with energies greater than the ankle energy at $\approx
5\times 10^{18}$ eV, are probably extragalactic protons in view of
their large gyroradii, lack of enhancements of arrival directions
toward the galactic plane, the observed flattening of the cosmic-ray
spectrum at the ankle (which could result from photopair energy-loss
processes on cosmic-ray protons \cite{wda04,bgg04}), and suggestions of a
composition change from heavy to light nuclear composition above
$\approx 10^{17.4}$ eV \cite{bir93} (see \cite{nw00} for a review).

Because high-energy protons will radiate $\gamma$-ray photons
from hadronic and leptonic processes at the site where they are
produced, the most likely candidate sources of UHECRs are
extragalactic nonthermal $\gamma$-ray emitters, specifically GRBs and
blazar AGNs.  Models for these sources involve jets of relativistic
plasma that are ejected by accreting black holes. The paucity of
blazar sources within tens of Mpc that could account for the near
isotropy of cosmic rays at energies $\gtrsim 5\times 10^{19}$ eV
suggest that GRBs are the sources of UHECRs. GRBs in the Galaxy will also
accelerate HECRs, and could form most of the cosmic rays with 
energies between $\approx 10^{14}$ eV and $\approx 5\times
10^{17}$ eV.

Several types of observations can test this hypothesis. Detection of
high-energy neutrinos from a GRB, which would requires an ultra-relativistic
hadronic component that is much more powerful than the nonthermal
electron component that produces the hard X-ray and soft $\gamma$-ray
emissions from GRBs \cite{da03}, is predicted by this model. Another
prediction is the detection of hadronic emission components in the
spectra of GRBs.  The unusual $\gamma$-ray emission component observed
\cite{gon03} in GRB 941017, which could be caused by
ultra-relativistic hadron acceleration and photopion processes in the
inner jets of GRBs \cite{da04}, could be the first example of such a
signature.

A third observation that would implicate GRBs as the sources of HECRs
is the detection of high-energy neutron $\beta$-decay halos around
star-forming galaxies \cite{der02}, including $\beta$-decay emission
from GRBs in the Milky Way \cite{ikm04}. A fourth observation is the
absence of $\gtrsim 10^{18}$ eV cosmic-ray {\it point-source}
excesses, such as the source claimed to be detected with the SUGAR
array. This is opposite to the conclusion arrived by \cite{bie04}, as
explained below, and will soon be tested with the Auger experiment,
which is now taking data.

\section{Complete Model for Cosmic Ray Origin}

We have recently proposed a model for HECRs from galactic and
extragalactic GRBs \cite{wda04}, building on previous suggestions that
UHECRs are accelerated by extragalactic GRBs \cite{vie95,wax95,mu96}.
In our model, relativistic outflows in GRBs are assumed to inject
power-law distributions of cosmic rays to the highest ($\gtrsim
10^{20}$ eV) energies.  A diffusive propagation model for HECRs from a
single recent GRB within $\approx 1$~kpc from Earth that took place
within the last 0.5 million years can explain the KASCADE data for the
CR ion spectra near and above the knee.  The CR spectrum at energies
 $\gtrsim 10^{18}$~eV is fit with CRs from extragalactic
GRBs.

UHECRs produced by extragalactic GRBs lose energy from momentum
redshifting and photo-pair and photo-pion production on the CMBR
during propagation through a flat $\Lambda$CDM universe (we take
$\Omega_{matter}=0.3$ and $\Omega_{\Lambda}=0.7$ in our calculations).
Attenuation produces features in the UHECR flux at characteristic
energies $\sim 4\times 10^{18}$~eV and $\sim 5\times 10^{19}$~eV due
to photo-pair and photo-pion energy-loss processes, respectively, from
distant sources at $z\gtrsim 1$.

The GRB rate-density evolution is assumed to follow the SFR history
derived from the blue and UV luminosity density of distant galaxies.
To accommodate uncertainty in the SFR evolution, we take two models,
one based on optical/UV measurements without extinction corrections
(lower SFR), and the other with extinction corrections (upper SFR).
The upper SFR is roughly a factor of 3(10) greater than the lower SFR
at red-shift $z=1(2).$ For $\gtrsim 10^{19}$~eV CRs, both evolution
models give the same flux, but the upper SFR contributes a factor
$\sim 3$ more CR flux over the lower SFR at energies $\lesssim
10^{18}$~eV.

\begin{figure}
\includegraphics[scale=0.5]{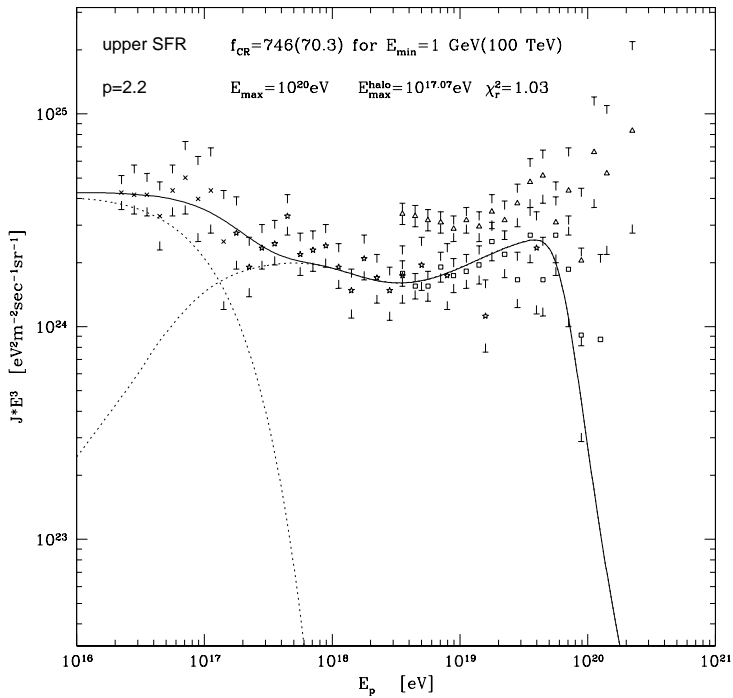}     
\caption{Best fit to the KASCADE (crosses), HiRes-I Monocular (squares), 
and HiRes-II Monocular (stars) data assuming a spectral cutoff at the
source of $E_{max}=10^{20}$~eV and using the upper limit to the SFR
evolution.  We also show the AGASA data (triangles), but do not
include these in our fits (see \cite{wda04} for references to the
data).  The cutoff energy for the galactic-halo component is
$E_{max}^{halo}=10^{17.07}$~eV and $\chi^{2}_r =1.03.$ The fit
implies that the transition from Galactic to extragalactic CRs occurs
near the second knee and that the ankle is associated with photo-pair
production. }
\end{figure}

The combined KASCADE, HiRes-I and HiRes-II Monocular data between
$\approx 2\times 10^{16}$~eV to $3\times 10^{20}$~eV are fit in Fig.\
1 which shows our best case, with cosmic-ray number injection index
$p=2.2$, $E_{max}=10^{20}$~eV, and the upper SFR.  A $p = 2.0$
spectrum provides a worse fit than the $p = 2.2$ case, although the CR
energy demand is less in this case because CRs are injected equally
per unit decade in particle energy.  The transition between galactic
and extragalactic CRs is found in the vicinity of the second knee at
$\approx 10^{17.6}$~eV, consistent with a heavy-to-light composition
change \cite{bir93}.  The ankle, at $\approx 10^{18.5}$~eV, is
interpreted as a suppression from photo-pair losses, analogous to the
GZK suppression.

\section{High-Energy Neutrinos} 

By normalizing the energy injection rate to that required to produce
the CR flux from extragalactic sources observed locally, we determine
the amount of energy a typical GRB must release in the form of
nonthermal hadrons. Our results imply that GRB blast waves are
baryon-loaded by a factor $f_{CR}\gtrsim 60$ compared to the primary
electron energy that is inferred from the fluxes of hard X-rays and
soft $\gamma$ rays measured from GRBs.

In Fig.\ 2 we show the neutrino fluences expected in the collapsar GRB
scenario from a burst with photon fluence $\Phi_{rad} = 3\times
10^{-4}\,\rm erg\, cm^{-2}$. The neutrino fluences are calculated for
3 values of the Doppler factor $\delta$ from a GRB at redshift $z = 1$
(we take a Hubble constant of 65 km s$^{-1}$ Mpc$^{-1}$ in the
calcuations). In order to demonstrate the dependence of the neutrino
fluxes on $\delta$, we consider 3 values of $\delta$ and set $f_{CR}
=20$ in this calculation. We assume that the prompt emission is
contributed by $N_{spk} = 50$ spikes with characteristic timescales
$t_{spk} \simeq 1 \rm \, s$ each, which defines the characteristic
size (in the proper frame) of the emitting region associated with each
individual spike through the relation $R_{spk}^\prime \simeq t_{spk}
\delta /(1+z)$.

The numbers of muon neutrinos that would be detected from a single
GRB with these parameters with IceCube for
$\delta = 100,\, 200$ and 300 are $N_\nu = 1.32,\, 0.105 $ and 0.016,
respectively. We should note, however, that for the assumed value of
$f_{CR}$, the calculated total fluence of neutrinos (both $\nu_\mu$
and $\nu_e$) produced when $\delta = 100$ is $\Phi_{\nu ,tot} =7.2
\times 10^{-4}$ erg cm$^{-2}$, i.e., by a factor $7.2/3 = 2.4$ {\it
larger} than the assumed radiation fluence.  This means that the
maximum value of the baryon loading that could be allowed if the
high-energy radiation fluence is less than the X/$\gamma$ fluence for
this particular case should be about 8 -- 10, instead of 20, in order
that the hadronic cascade $\gamma$-ray flux is guaranteed not to
exceed the measured photon flux. Consequently, the expected number of
neutrinos for $\delta=100$ should be reduced to $\simeq 0.6$. On the
other hand, the neutrino fluence for the case $\delta =200 (300)$ is
equal to $\Phi_{\nu,tot} =1.4 \times 10^{-4} (3 \times 10^{-5}) \,\rm
erg\, cm^{-2}$, so this accommodates an increased baryon-loading from
$\lesssim 20$ up to $f_{CR}\simeq 45 (200)$, with the expected number
of neutrinos observed by IceCube being $N_{\nu ,corr} \simeq 0.23
(0.16)$.  If the radiation fluence at MeV -- GeV energies is allowed
to exceed the X/$\gamma$ fluence by an order of magnitude, a
possibility that GLAST will resolve, then the expected number of
detected neutrinos could be increased correspondingly.

\begin{figure}
\includegraphics[scale=0.5]{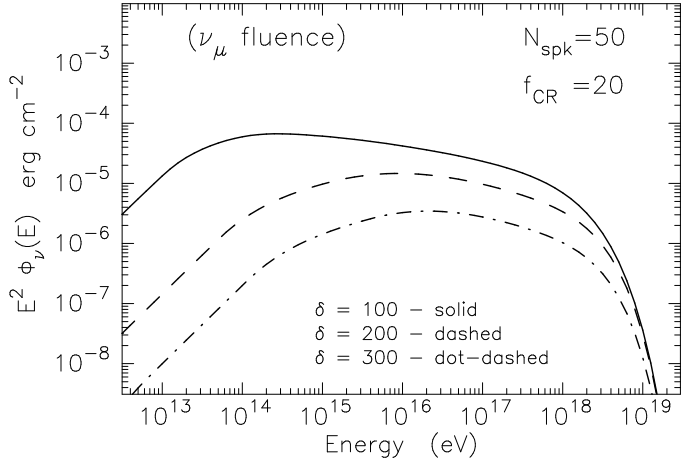}     
\caption{ The fluences of muon neutrinos calculated 
for a GRB at $z = 1$ with X-ray/MeV $\gamma$-ray fluence of $3\times
10^{-4}$ ergs cm$^{-2}$ and Doppler factors $\delta = 100$, 200 and
300, and a nonthermal baryon-loading factor $f_{CR}= 20$.  }
\end{figure}

For the large baryon load required for the proposed model of HECRs,
calculations show that 100 TeV -- 100 PeV neutrinos could be detected
several times per year from all GRBs with kilometer-scale neutrino
detectors such as IceCube \cite{da03,wda04}.  Detection of even 1 or 2
neutrinos from GRBs with IceCube or a northern hemisphere neutrino
detector will provide compelling support for this scenario
for the origin of high-energy and UHE cosmic rays.

\section{Hadronic Emission Components in GRB Spectra}

Based on joint analysis of BATSE Large Area Detector and the EGRET
Total Absorption Shower Counter data, Gonz\'alez et al.\
\cite{gon03} reported the detection of an anomalous MeV emission
component in the spectrum of GRB 941017 that decays more slowly than
the prompt emission detected with the LAD in the $\approx 50$ keV -- 1
MeV range. The multi-MeV component lasts for $\gtrsim 200$ seconds
(the $t_{90}$ duration of the lower-energy prompt component is 77
sec), and is detected with the BATSE LAD near 1 MeV and with the EGRET
TASC between $\approx 1$ and 200 MeV. The spectrum is very hard, with
a photon number flux $\phi(\epsilon_\gamma)\propto
\epsilon_\gamma^{-1}$, where $\epsilon_\gamma $ is the observed photon
energy.

This component is not predicted or easily explained within the
standard leptonic model for GRB blast waves, though it possibly could
be related to Comptonization of reverse-shock emission by the forward
shock electrons \cite{gg03}, including self-absorbed reverse-shock
optical synchrotron radiation \cite{pw04}. Another possibility is
that hadronic acceleration in GRB blast waves could be responsible for
this component.

We have argued \cite{da04} that this component could be a consequence
of the acceleration of hadrons at the relativistic shocks of GRBs. A
pair-photon cascade initiated by photohadronic processes between
high-energy hadrons accelerated in the GRB blast wave and the internal
synchrotron radiation field produces an emission component that
appears during the prompt phase, as shown in Fig.\ 3. Photomeson
interactions in the relativistic blast wave also produce a beam of UHE
neutrons, as proposed for blazar jets \cite{ad03}. Subsequent
photopion production of these neutrons with photons outside the blast
wave will produce a directed hyper-relativistic electron-positron beam
in the process of charged pion decay and the conversion of high-energy
photons formed in $\pi^0$ decay. These energetic leptons produce a
synchrotron spectrum in the radiation reaction-limited regime
extending to $\gtrsim$ GeV energies, with properties in the 1 -- 200
MeV range similar to that measured from GRB 941017. GRBs displaying
these anomalous $\gamma$-ray components are most likely to be detected
as sources of high-energy neutrinos \cite{gue04}.

\begin{figure}
\includegraphics[scale=0.5]{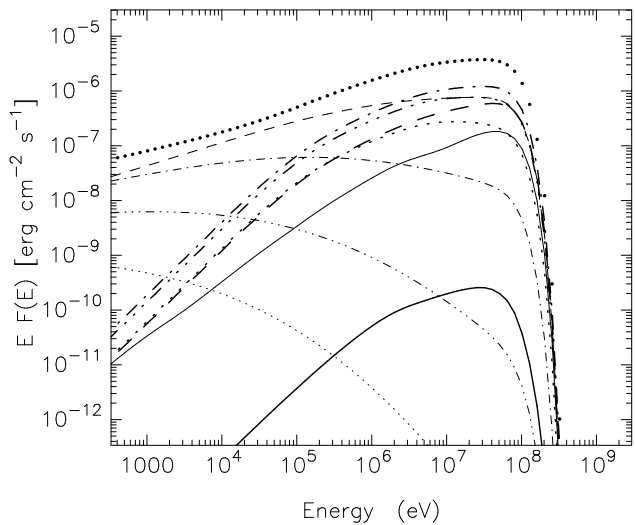}     
\caption{Photon energy fluence from an electromagnetic cascade initiated
by photopion secondaries in a model GRB, with parameters given in
Ref.\ \cite{da04}. Five generations of Compton (heavy curves) and
synchrotron (light curves) are shown. The first through fifth
generations are given by solid, dashed, dot-dashed,
dot-triple--dashed, and dotted curves, respectively. The total cascade
radiation spectrum is given by the upper bold dotted curve. }
\end{figure}

\section{Neutron $\beta$-Decay Halos}

If GRBs are the sources of HECRs, then high-energy neutrons will be
formed at the burst site through photo-pion processes and, being
neutral, can escape to intergalactic space. Indeed, it is possible
that charged ultra-high energy protons and ions are trapped by ambient
magnetic field and lose energy adiabatically near the acceleration
site, so that UHECRs are made primarily of neutron-decay protons.

The decay of the neutrons far from the GRB through the process $n
\rightarrow p + e^- +\bar \nu_e$ leads to $\beta$-decay electrons that
make synchrotron and Compton radiation. The best prospect for
discovering neutron-decay halos is to search for diffuse optical
synchrotron halos surrounding field galaxies that display active star
formation \cite{der02}. GRBs that have recently taken place in our
Galaxy will produce Compton radiation at TeV energies that could be
detected by the {\it HESS} and {\it VERITAS} imaging air Cherenkov
telescopes \cite{ikm04}. The emission of nonthermal synchrotron and
Compton radiation from photopion processes by UHECRs traveling through
intergalactic space will also produce a nonthermal component of the
diffuse radiation background, which can be used to determine the
magnetic field of intergalactic space.

\section{GRBs in the Galaxy}

Because the Milky Way is actively making young high-mass stars, GRBs
will also occur in our Galaxy. The rate of GRBs in the Milky Way is
very uncertain because of lack of precise knowledge about the opening
angle of GRB jets, but could be as frequent as once every 10,000
years. Over the age of the Galaxy, there is a good chance that a
nearby powerful GRB with a jet oriented towards Earth could have
lethal consequences for life. It has recently been argued \cite{mel04}
that such an event contributed to the Ordovician extinction event 440
Myrs ago. 

To assess cosmic ray transport from a GRB, we \cite{dh05} developed a
3D propagation model to simulate the sequence of irradiation events
that occurs when a GRB jet is pointed towards Earth. The cosmic rays
move in response to a large-scale magnetic field that traces the
spiral arm structure of the Galaxy, and they diffuse through
pitch-angle scattering with magnetic turbulence. The magnetic
field of the Galaxy is modeled as a bi-symmetric spiral for the
Galaxy's disk, and a dipole magnetic field for the Galaxy's halo
\cite{alv02}. The evolution of the particle momentum is found by
solving the Lorentz force equation. A Monte Carlo simulation of
pitch-angle scattering and diffusion was developed that takes into
account the energy dependence of the cosmic-ray mean free path.

Fig. 4 displays the cosmic-ray halo that surrounds a GRB source 14,000
years after the GRB event. The geometry of the system is modeled
by twin radial jets of cosmic rays with a jet-opening half-angle of
0.1 radian. A conical shell forms as a result of protons and
neutron-decay protons with energies $\gtrsim 10^{18}$ eV.

\begin{figure}
\includegraphics[scale=0.5]{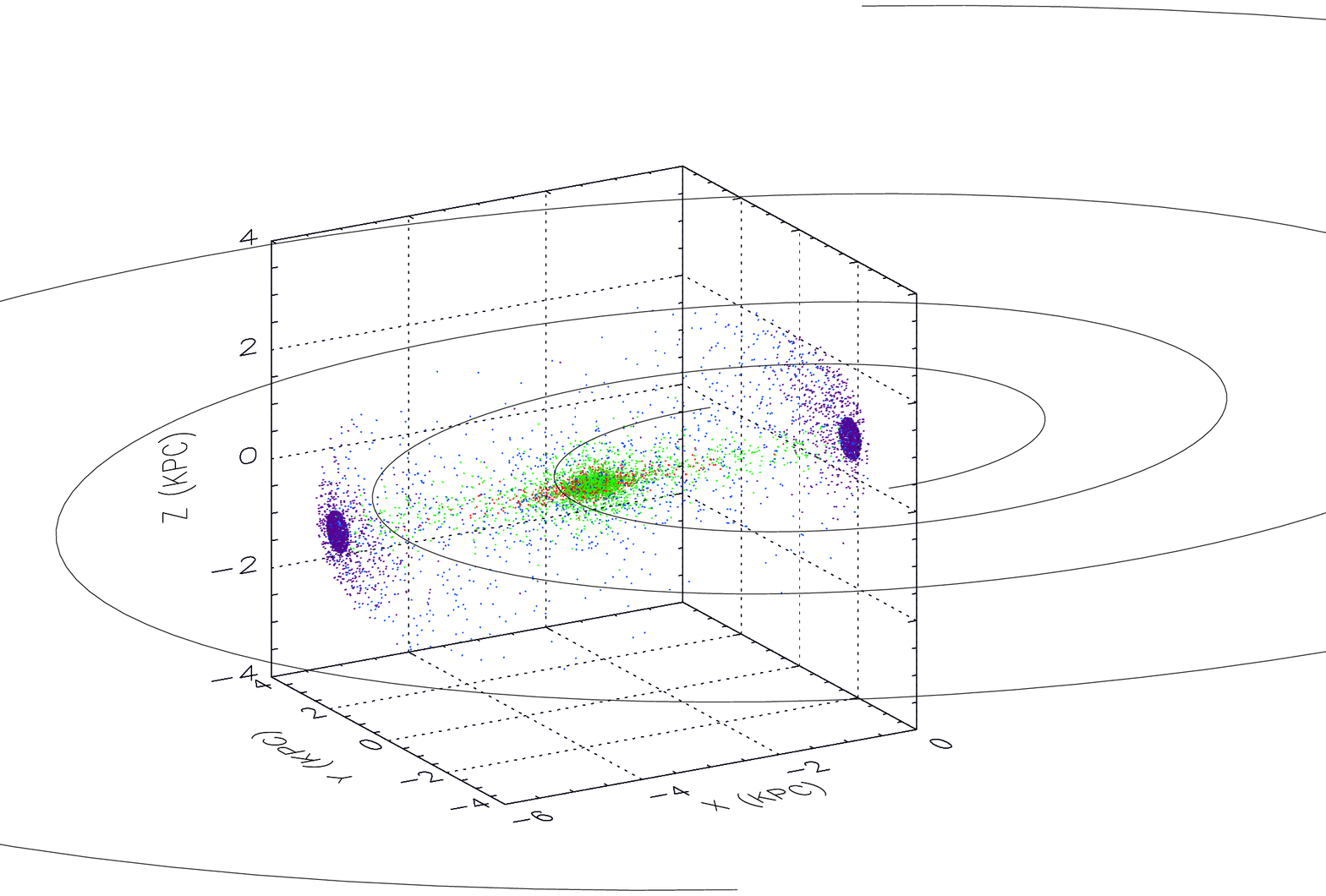}     
\caption{Cosmic-ray halo formed 14,000 years after a GRB, with 
radial jets oriented along the galactic plane, that took place at 3
kpc from the center of the Galaxy. For clarity, equal numbers of
cosmic rays are injected per decade between $10^{16}$ -- $10^{20}$
eV. }
\end{figure}

The hypothesis \cite{bie04} that the $\gtrsim 10^{18}$ eV cosmic-ray
excesses detected with the AGASA (Akeno Giant Air Shower Array) and
SUGAR arrays \cite{hay99,bel01} are cosmic-ray neutrons from a GRB is
not, however, supported by our simulations.  The relativistic blast
waves in a GRB accelerate the highest energy cosmic rays over
timescales of weeks or less \cite{zm04}, and the high-energy neutrons
therefore arrive on this same timescale. Cosmic ray protons with
energies $\sim 10^{19}$ eV are delayed over a timescale $\approx
10000\theta^3/B_{\mu{\rm G}}$ years from a source at the distance of
the Galactic Center. For the SUGAR excess, which is coincident on the
sub-degree angular scale with a point source, a GRB would have to take
place within weeks of the observation for cosmic-ray protons to
maintain their direction to the source.  Including the requirement
that the GRB jet was also pointed towards Earth means that an
impulsive GRB origin is excluded because such an event is so
improbable.  The greater ($\approx 10^\circ$) extent of the AGASA
excess does not conclusively exclude a GRB origin, but here the
diffuse excess could simply reflect the greater pathlength for
cosmic-ray proton collisions with spiral arm gas along the Cygnus arm.

Because the SUGAR point source does not admit an impulsive GRB
solution, only cosmic rays from a persistent source, such as a
microquasar, could make such an excess. The hypothesis
\cite{der02,wda04} that GRBs are sources of $\gtrsim 10^{14}$ eV 
cosmic rays is therefore incompatible with such a source. Our
cosmic-ray origin hypothesis will soon be tested by results from the
Auger Observatory\footnote{www.auger.org} to confirm or refute the
existence of the SUGAR source. If the source is real, then our
GRB/cosmic-ray model is incomplete.

\section{Conclusions}

We have summarized a complete model \cite{wda04} for HECRs from
galactic and extragalactic GRBs. Our interpretation of the HECR
spectrum requires that GRBs are hadronically dominated, which is
necessary \cite{da03} for neutrinos from GRBs to be detectable with a
km-scale neutrino telescope. Detection of high-energy neutrinos from
GRBs will provide strong support for this model.

Searches for anomalous $\gamma$-ray emission components with {\it
GLAST}, as already observed form GRB 941017 with BATSE and EGRET, will give
additional support for cosmic-ray acceleration in GRBs, provided that
these components can be attributed to hadronic cascade radiation
\cite{da04}. Observations of neutron $\beta$-decay radiation 
from recent GRBs in the Galaxy \cite{ikm04} and around galaxies that
host GRBs \cite{der02} provide further tests of the hypothesis that
high-energy cosmic rays are accelerated by GRBs.

The ground-based Auger experiment, which will measure air showers and
Nitrogen fluorescence from the UHECR interactions in the atmosphere,
is now taking data. This experiment will greatly expand our knowledge
of high-energy cosmic rays, for example, by measuring with high
quality statistics and checks on calibration the spectrum of $\gtrsim
10^{20}$ eV cosmic rays which produce high-energy neutrinos at their
source and as they propagate through intergalactic space. Detection of
$\approx 10^{18}$ eV cosmic-ray point sources is incompatible with 
an origin from impulsive GRBs, so that if the SUGAR source is confirmed,
this cosmic-ray origin model is incomplete. 

\acknowledgments
I am pleased to acknowledge my collaboration with Armen Atoyan, Jeremy
Holmes, and Stuart Wick.  This work was supported by the Office of
Naval Research and a NASA {\it Gamma ray Large Area Space Telescope}
grant.

\end{document}